# Transport Spin Polarization of High-Curie Temperature MnBi Films

P. Kharel[1, 2], P. Thapa[3], P. Lukashev[1, 2], R. F. Sabirianov[4, 2], E. Y. Tsymbal [1, 2], D. J. Sellmyer[1, 2], and B. Nadgorny[3]

[1]*Department of Physics and Astronomy, University of Nebraska, Lincoln, NE, 68588*

[2]*Nebraska Center for Materials and Nanoscience, University of Nebraska, Lincoln, NE, 68588*

[3]*Department of Physics and Astronomy, Wayne State University, Detroit, MI, 48202*

[4]*Department of Physics and Astronomy, University of Nebraska, Omaha, NE, 68182*

## Abstract

We report on the study of the structural, magnetic and transport properties of highly textured MnBi films with the Curie temperature of 628K. In addition to detailed measurements of resistivity and magnetization, we measure transport spin polarization of MnBi by Andreev reflection spectroscopy and perform fully relativistic band structure calculations of MnBi. A spin polarization from $51\pm1$ to $63\pm1\%$ is observed, consistent with the calculations and with an observation of a large magnetoresistance in MnBi contacts. The band structure calculations indicate that, in spite of almost identical densities of states at the Fermi energy, the large disparity in the Fermi velocities leads to high transport spin polarization of MnBi. The correlation between the values of magnetization and spin polarization is discussed.

## I. INTRODUCTION

Successful implementation of many novel concepts and devices in spintronics is largely dependent on our ability to controllably generate and inject electronic spins, preferably at room temperature[1], which require spin injectors to combine high Curie temperature with reasonably high conductivity. Unlike all-metal devices, where efficient electrical spin injection has been demonstrated[2], spin injection from ferromagnetic metals into semiconductors proved to be more challenging, partly because of the low interface resistance[3]. This problem may be circumvented by spin injection from 100% spin polarized, half-metallic contacts, tunnel contacts, or semiconductor contacts. While a number of promising magnetic semiconducting systems, such as (Ga,Mn)As, for example, have been investigated[4], their relatively low Curie temperatures make practical implementation of these materials difficult. Doping some of the magnetic oxides with magnetic ions represents another approach; however, the progress in this area has been slow due in part to persisting reproducibility problems.[5]

The interest in ferromagnetic MnBi stems from its high Curie temperature, which is well above room temperature[6], high coercivity with a rectangular hysteresis loop[7], large perpendicular room-temperature anisotropy in thin films[8] that can be used as spin injectors for spin lasers and spin emitting diodes[9], and an extraordinarily large Kerr rotation[10]. The ferromagnetic phase in the NiAs structure is the most stable at room temperature, undergoing a coupled structural and magnetic phase transition at 628K. These unusual magnetic and magneto-optical properties have been the main motivation for the intensive studies on the various properties of this material.[11] Recently it has been predicted that MnBi in the hypothetical zinc blende structure is fully half-metallic.[12,13,14] The experimental implementation of the zinc blende MnBi may be quite challenging – not only because it is difficult to grow MnBi epitaxially, but also because the zinc blende phase may be metastable. On the other hand, MnBi in the NiAs structure can be fabricated, is ferromagnetic up to 628 K, and is a fairly good conductor at room temperature. Moreover, the properties of MnBi interface may be controlled by the addition of Bi, which shows a semimetal-semiconductor transition at small thicknesses.[15]

From this perspective it is particularly important to measure the transport spin polarization of MnBi in the NiAs structure, which is also relevant to the understanding of MnBi junctions that show a large magnetoresistance (70% at room temperature).[16]

The question of maximizing the value of the transport spin polarization $P_T$ is often discussed in the context of possible correlation of $P_T$ with the value of magnetization $M$, or the average atomic magnetic moment of a ferromagnet. Experimentally, while the linear relationship between $P_T$ and $M$ has been reported[17], in many other cases no direct relationship between the two quantities has been observed.[18,19,20] As $P_T$ is associated with the electronic states near the Fermi energy and the respective Fermi velocities, whereas the magnetic moment is associated with the algebraic sum of occupancies of all majority and minority spin states, there is no reason for these quantities to be related. Thus the determination of whether or not the link between the two quantities exists in a concrete materials system has to be made independently.

Here we report the Point Contact Andreev Reflection (PCAR)[21,22] measurements of the transport spin polarization, $P_T$ of MnBi thin films in the NiAs crystallographic structure. We find a relatively large spin polarization of up to 63%, consistent with our density functional calculations and an observation of a large magnetoresistance in MnBi contacts.[16] We also report a correlation between the values of the saturation magnetization and the transport spin polarization.

## II. EXPERIMENTAL RESULTS

MnBi thin films were prepared by sequential evaporation of Bi and Mn onto a glass substrate using an e-beam evaporator with subsequent *in situ* annealing of bi-layers immediately after the deposition. High quality MnBi thin films can be grown by this method, if the Mn to Bi atomic ratio of 55 to 45 is maintained during deposition.[23] Here we will present the data taken on four samples with the thicknesses from approximately 32 nm (samples A, C, and D) to 47 nm (sample B). Two samples (A and C) were deposited at room temperature and annealed for one hour at $410^{\circ}$C and $400^{\circ}$C respectively; the other two samples (B and D) were deposited at $125^{\circ}$C and annealed at $350^{\circ}$C for one and a half and one hour respectively. All of the samples were single phase MnBi highly textured polycrystalline films, with a hexagonal NiAs crystal structure, although small traces of elemental Bi have been detected (see Fig.1).

Depending on the experiment, several generally different definitions of spin polarization has been introduced.[24] $P_T$ is defined as $P_{Nv} = \dfrac{\langle Nv \rangle_\uparrow - \langle Nv \rangle_\downarrow}{\langle Nv \rangle_\uparrow + \langle Nv \rangle_\downarrow}$, or as $P_{Nv^2} = \dfrac{\langle Nv^2 \rangle_\uparrow - \langle Nv^2 \rangle_\downarrow}{\langle Nv^2 \rangle_\uparrow + \langle Nv^2 \rangle_\downarrow}$ in the case of the ballistic and diffusive regimes respectively, where $v$ is the Fermi velocity, and $N$ is the spin-projected densities of states (DOS) for majority (↑) and minority (↓) spins respectively. In the ballistic regime only one component of velocity predominantly enters the averaging. For all of the point contact measurements described here electrochemically etched Nb tips were used[25]. The differential conductance $dI/dV$ was obtained by a four-probe technique with standard ac lock-in detection at a frequency of approximately 2 kHz. The details of the experimental techniques and the data analysis can be found in Ref. 26. Since, as we will show below, all the contacts are largely in the ballistic regime, we used the modified Blonder-Tinkham-Klapwijk (BTK)[25] model[27] in the ballistic regime to analyze the data. The typical conductance curves for samples A, B, C, and D are shown in Fig. 2a. To account for possible empirical $Z^2$ dependence of the spin polarization values on a scattering parameter $Z$ at the F/S interface, often encountered in the PCAR measurements[28], we plotted $P(Z)$ dependencies for the respective samples in Fig. 2b taking the extrapolation of the least square fit to the case of transparent interface ($Z = 0$) to obtain the limiting values of $P_T$. This procedure resulted in spin polarizations of 63±0.8, 57.8±1.6, 54.2±2.4, and 51.7±1.1%, for samples A, B, C, and D respectively.

We find that the values of the spin polarization are correlated with the magnetic properties of MnBi films. Magnetic hysteresis curves show that the samples are highly anisotropic with the magnetization easy axis perpendicular to the sample plane, with very high values of uniaxial anisotropy constants $K_1$ and $K_2$, consistent with the previous reports.[23] While all of the samples show well defined, rectangular hysteresis loops in the out of plane geometry (see top left inset in Fig. 3), the magnetization and coercivity seem to be very sensitive to the sample preparation conditions. Specifically, the measured saturation magnetizations are 503, 485, 464 and 425 emu/cm$^3$ and coercivities are 8.4, 3.2, 7.9 and 5.4 kOe at 300 K for the samples A, B, C, and D respectively. As can be seen from Fig. 3 the experimental values of $P_T$ correlate with the values of the saturation magnetization of MnBi.[29] We attribute this behavior to magnetic disorder which may have adverse effects on the values of magnetization and spin polarization, as has been reported for $SrFeMoO_6$, for example.[30] We discuss this behavior below in view of our first-principles calculations.

All MnBi samples are metallic and exhibit a qualitatively similar temperature dependence of the resistivity $\rho$ from 2 to 300 K (see Fig. 4). The residual resistance ratio $\rho_{RT}/\rho_{4K}$ is almost the same (~ 8.5) for all the samples, with $\rho_{4K}$ ~ 15 $\mu\Omega$ cm. Surprisingly, we found that the low temperature (4 K< T < 30 K) resistivity of all the samples follow an anomalous power law, different from the $\rho \sim T^2$ expected for weakly ferromagnetic metals, due to a single magnon scattering mechanism[31]. The resistivity of our samples follow the $\rho \sim T^m$ power law with m between 2.9 and 3.6, similarly to what has been observed in some half-metallic films, such as $CrO_2$[32]. While it has been suggested that the $T^3$ power law may be related to the unconventional single magnon scattering mechanism in half metals due to the spin fluctuations at finite temperatures[33], our results on MnBi indicate that it cannot be considered a definitive test for half-metallicity.

### III. COMPUTATIONAL RESULTS

To interpret the measured values of spin polarization we have implemented electronic band structure calculations of bulk MnBi in the NiAs phase, using the tight-binding linear muffin-tin orbital (LMTO) method[34], within the local density approximation (LDA). We performed fully relativistic calculations, i.e. the scalar relativistic wave equation is solved. To explore the role of spin-orbit interaction (SO), we carried out the calculations both with and without taking spin orbit coupling into account. Somewhat surprisingly, we find that SO practically does not affect our results (see Figs. 5 and 6). While there is a slight band shift on the order of SO constant (Fig. 5), we found practically no difference in the total calculated DOS. Close inspection also shows that the inclusion of SO does not significantly change the dispersion relationships at the Fermi level (Fig. 6). Consequently the Fermi velocities - and thus the values of transport spin polarization would only be marginally affected by the inclusion of the SO coupling.

Fig. 7 (top panel, shaded region) shows that the total densities of states (DOS) at the Fermi energy are nearly equal (~ 0.45 states/cell eV) for majority- and minority-spin carriers, resulting in a vanishing spin polarization, $P_N = \dfrac{N_\uparrow - N_\downarrow}{N_\uparrow + N_\downarrow}$, where $N_\uparrow$ and $N_\downarrow$ are the majority- and minority-spin DOS (see bottom panel of the Fig. 5). The origin of the large $P_T$ measured in MnBi is due to the substantial spin asymmetry of the electronic bands near the Fermi energy. The close inspection of the dispersion of the minority and majority bands (see Fig. 8) indicates that the minority spin states have a lower Fermi velocity compared to the majority bands. Indeed, the calculated Fermi velocities $v_{\uparrow(\downarrow)}$ are $1.2\times10^6$ and $0.6\times10^6$ m/s for the majority and minority bands respectively (both are almost constant in the range $\pm 0.5$ eV around $E_F$). Thus, when the mobility of electrons is taken into account, a large $P_T$ is expected.[35,36,37] The definition of $P_T$ in the diffusive regime assumes that the relaxation time which enters the expression for the conductivity is spin-independent.[37] This may be qualitatively justified given the fact that the relaxation time is proportional to the DOS at the Fermi energy,[38] but the latter is nearly spin-independent according to our calculations. The calculations yield the spin polarization $P_{Nv} = 36\%$ and $P_{Nv^2} = 66\%$ assuming that the Fermi velocity is projected to the c-axis (perpendicular to the plane of the film). Both $P_{Nv}$ and $P_{Nv^2}$ are reduced for the velocity direction perpendicular to the c-axis, i.e. in the *ab*-plane ($P_{Nv} = 28\%$ and $P_{Nv^2} = 51\%$). This implies that lower values of spin polarization are expected for polycrystalline MnBi samples due to the strong anisotropy of transport properties of MnBi.

To examine the correlation between saturation magnetization and spin polarization we used the fixed-spin moment method[39]. The results shown in Fig. 3 (bottom inset) indicate an approximately linear relationship between $P_T$ and the magnetic moment. This behavior is the consequence of nearly linear variation of the exchange splitting of the spin bands with the magnetic moment. Experimentally, the variation of the saturation magnetization may be due to a different degree of structural disorder in our samples. As follows from our calculations[40], placing Mn atoms in the interstitial sites leads to the antiferromagnetic alignment of their moments with the moments of the Mn atoms in the regular sites. The magnetization of MnBi decreases together with the value of spin polarization, supporting the experimentally observed trend in our MnBi samples.

## IV. DISCUSSSION AND CONCLUSIONS

Using the measured value of the resistivity of MnBi (~ 15 μΩ cm at 4 K) and the calculated density of states, $N_\uparrow = 0.446$ and $N_\downarrow = 0.425$ states/eV cell, we estimate the mean free path for majority (↑) and minority (↓) carriers from the Ziman formula $\sigma_{\uparrow(\downarrow)} = 1/3 e^2 N_{(\uparrow\downarrow)} v_{F(\uparrow\downarrow)}^2 \tau$, $L_\uparrow \approx 20 nm, L_\downarrow \approx 10 nm$. Using Wexler's formula[41] $R_c \approx 4\rho L / 3\pi d^2 + \rho / 2d$ the contact size can be estimated from 5 – 15 nm depending on the contact resistance, $10\ \Omega < R_C < 100\ \Omega$, indicating that the transport is in the ballistic regime for majority carriers and in the intermediate regime ($d \sim L$) for minority carriers. While these estimates suggest that our conditions correspond to the ballistic regime, our experimental results yield a better agreement with the theoretical calculations in the diffusive ($P_{Nv^2} = 51$- 66%), rather than in the ballistic ($P_{Nv} = 28$ - 36%) limit. A possible explanation is that the spin polarization can often be very sensitive to the interface, and to the termination of electrodes.[42] In MnBi it is expected to be strongly dependent on the surface termination because of the substantial difference in the electronic DOS at the Fermi energy for Bi and Mn. We find that $P_T$ is enhanced assuming the Bi states control the magnitude of $P$ ($P_{Nv} = 55\%$ and $P_{Nv^2} = 76\%$ respectively).

In conclusion, we have investigated the structural, magnetic and transport properties of high Curie temperature MnBi films. A transport spin polarization was measured using the Point Contact Andreev Reflection technique and values up to 63% are obtained, consistent with observations of a large magnetoresistance in MnBi contacts and the results of band structure calculations. Our first-principles calculations indicate that, in spite of almost identical densities of states at the Fermi energy in the majority- and minority-spin bands, the large disparity in the Fermi velocities results in a high transport spin polarization of MnBi. Our experimental data and first-principles calculations show a nearly linear relationship between the values of $P_T$ and the magnetic moment (magnetization) of MnBi.

The work at University of Nebraska was supported by the NSF-MRSEC (Grant DMR-0820521), the DOE grant DE-FG02-04ER46152, the Nebraska Research Initiative, and NCMN. The work at Wayne State was supported by the NSF CAREER ECS-0239058 and the ONR grant N00014-06-1-0616.

FIGURE CAPTIONS:

**Fig. 1**. (Color Online). XRD spectra of MnBi film (samples A and C). Strong diffraction peaks from (002) and (004) planes show preferred c-axis orientation of the films.

**Fig. 2**. (Color Online). Top: Examples of normalized conductance curves for samples A, B, C, and D. Sample A: contact resistance $R_C = 36.1\ \Omega$, the fitting parameters $Z = 0.15$ and $P = 61\%$; Sample B: $R_C = 50.5\ \Omega$, $Z = 0.28$ and P = 52%. Sample C: $R_C = 28.8\ \Omega$, $Z = 0.46$, P = 49 %; Sample D: $R_C = 17.9\ \Omega$, $Z = 0.43$ and $P = 42.4\%$; The BCS gap of niobium $\Delta = 1.5$ meV is used. (b) Bottom: $P\ (Z)$ dependence for samples A, B, C, and D respectively. The size of the data points corresponds to the error bars in $Z$ and $P \sim 0.02$.

**Fig. 3**. (Color Online). Spin polarization $P$ vs. saturation magnetization $M_s$ for samples A, B, C, and D. The straight line is constrained to go through the origin. Top inset: $M\ (H)$ loop for sample C in the magnetic field parallel and perpendicular to the c-axis. Bottom inset: calculated spin polarization vs. magnetic moment per MnBi unit cell.

**Fig. 4**. (Color Online). Resistivity of MnBi film as a function of temperature (samples A and C), showing the metallic behavior with the residual resistivity of 15 μΩcm. Inset: The power law dependence $\rho \sim T^m$ at low temperatures (below 30 K) with m = 2.9.

**Fig. 5.** (Color Online). Comparison between the DOS without SO (solid black line) and with SO (dashed blue line).

**Fig. 6.** Energy bands for majority spin channel without SO (leftmost panel), minority spin channels without SO (middle panel), both majority and minority spin channels with SO (rightmost panel).

Fig. 7. (Color Online). Dispersion of the minority and majority bands near Fermi level. Blue spheres – minority band; red squares – majority band.

**Fig. 8**. (Color Online). Density functional calculations: top panel – total DOS for majority and minority carriers (shaded region), $\langle Nv \rangle_{\uparrow(\downarrow)}$ (blue solid line), $\langle Nv^2 \rangle_{\uparrow(\downarrow)}$ (dashed red line); bottom panel – $P$ near the Fermi energy for $P_N$ (DOS) (black solid line crossing zero at 0 eV), $P_{Nv}$ (solid blue line), and $P_{Nv^2}$ (dashed red line) in the direction of the c-axis. Inclusion of spin - orbit coupling (from fully relativistic calculations) practically does not affect the calculated DOS as seen in Fig. 5

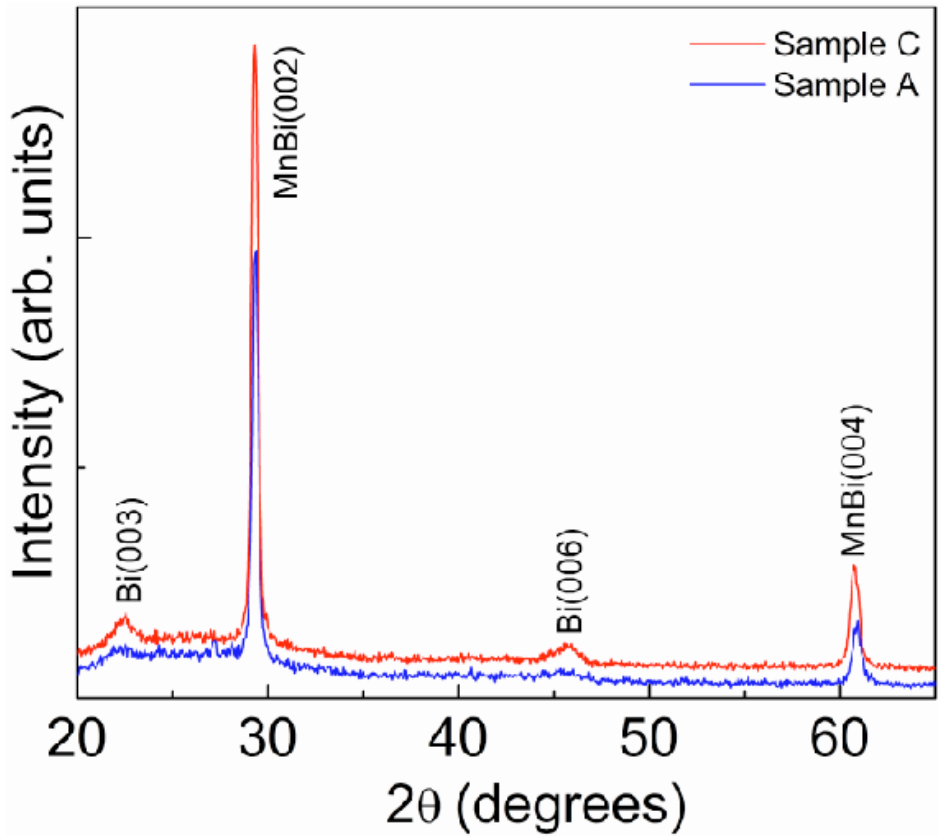

Fig. 1

Kharel et al,

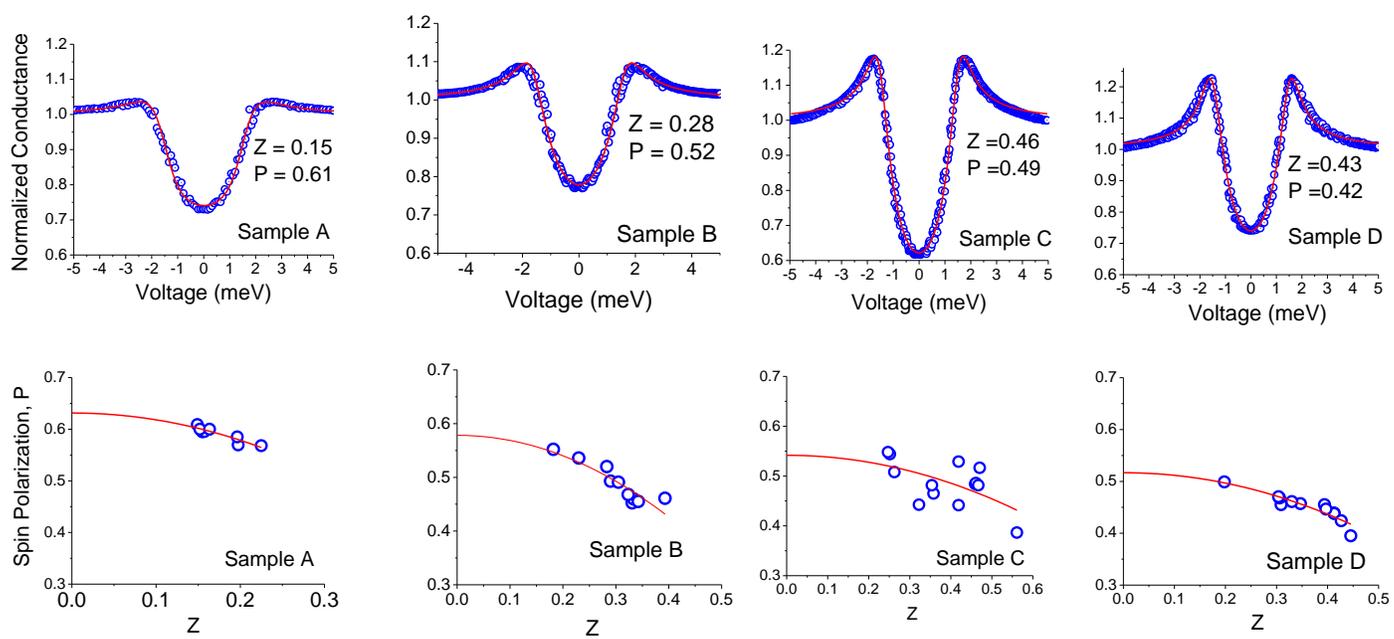

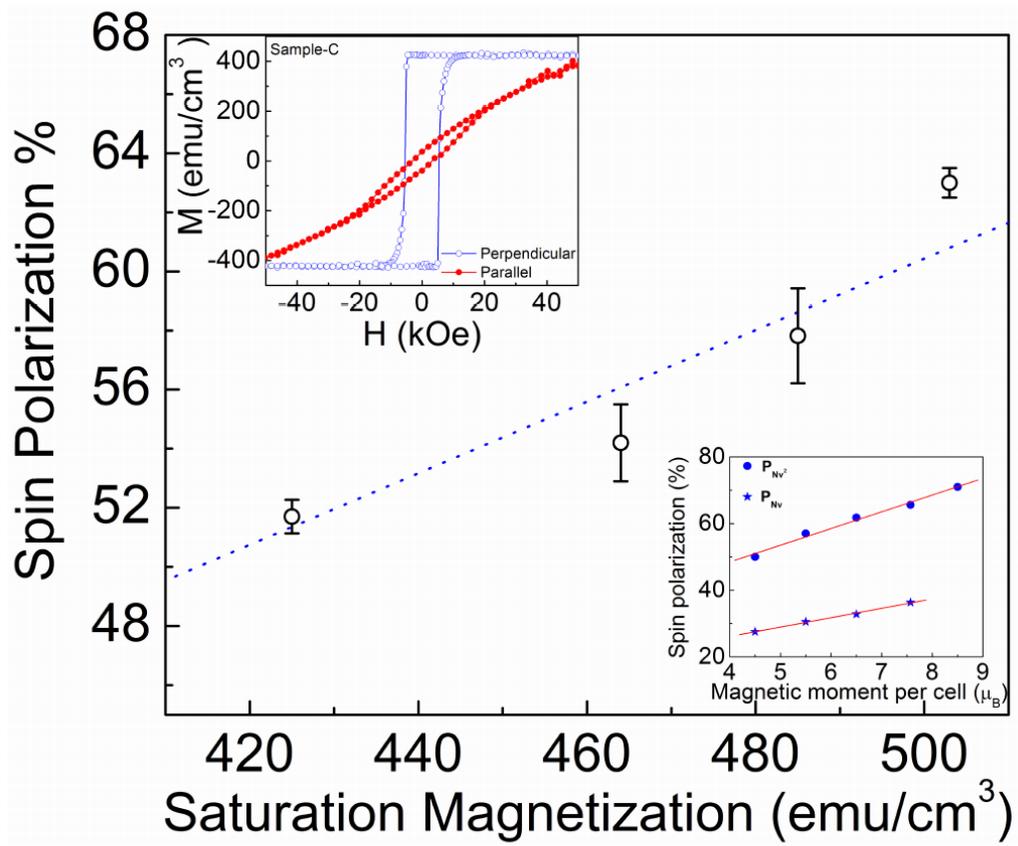

Kharel et al,

Fig. 3

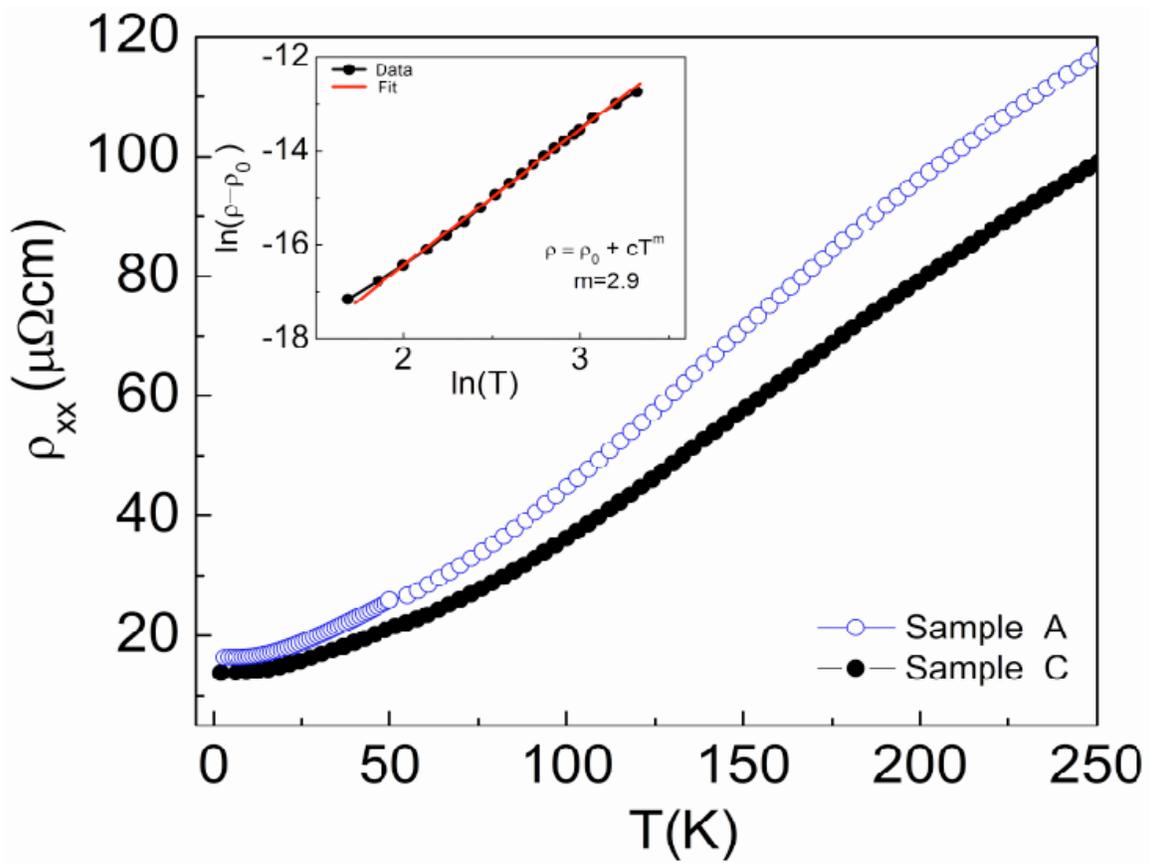

Kharel et al,

Fig. 4

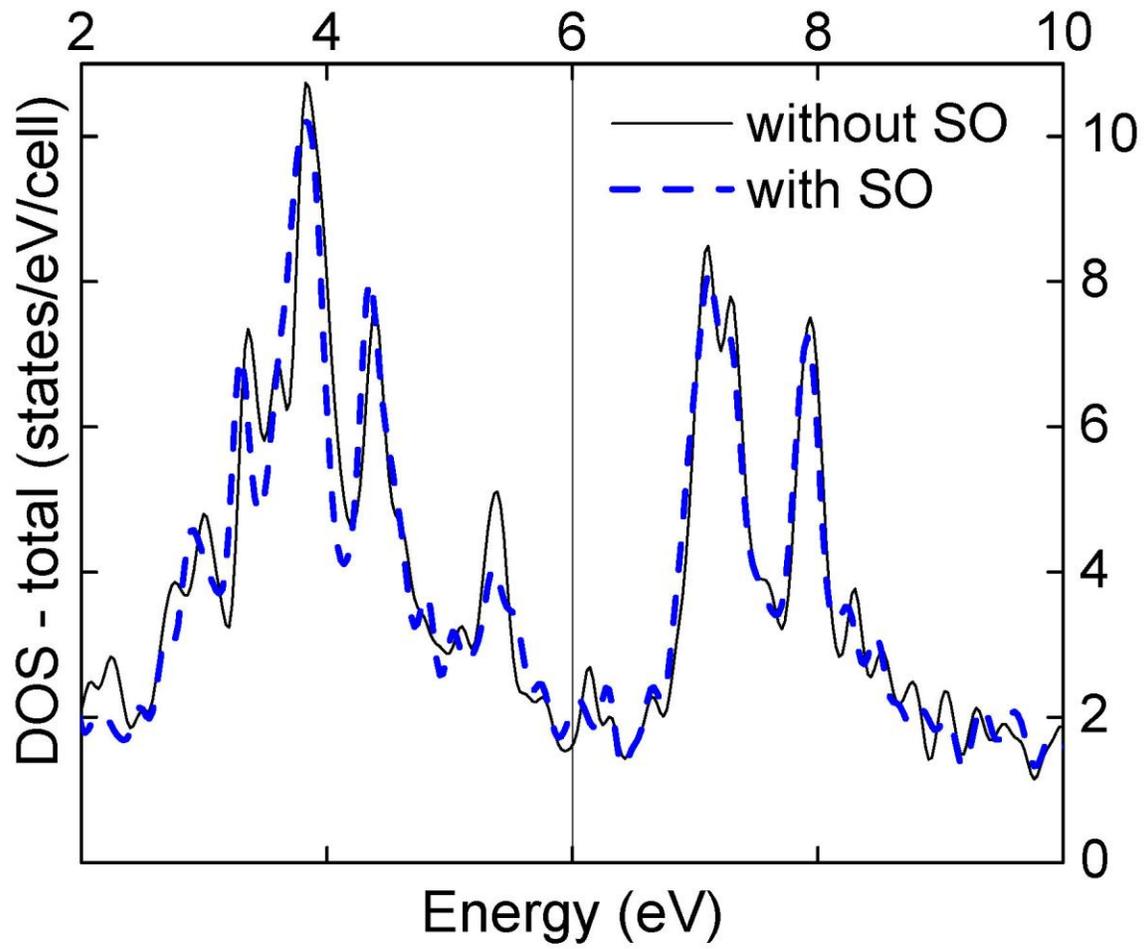

Kharel et al,

Fig. 5

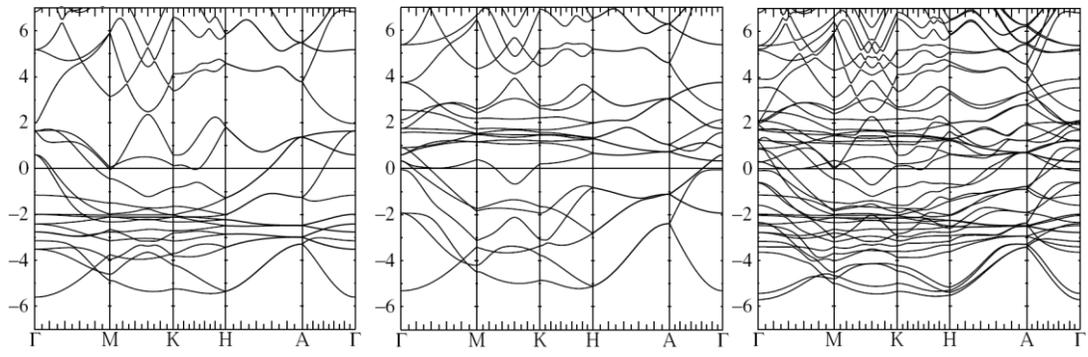

Kharel et al,

Fig. 6

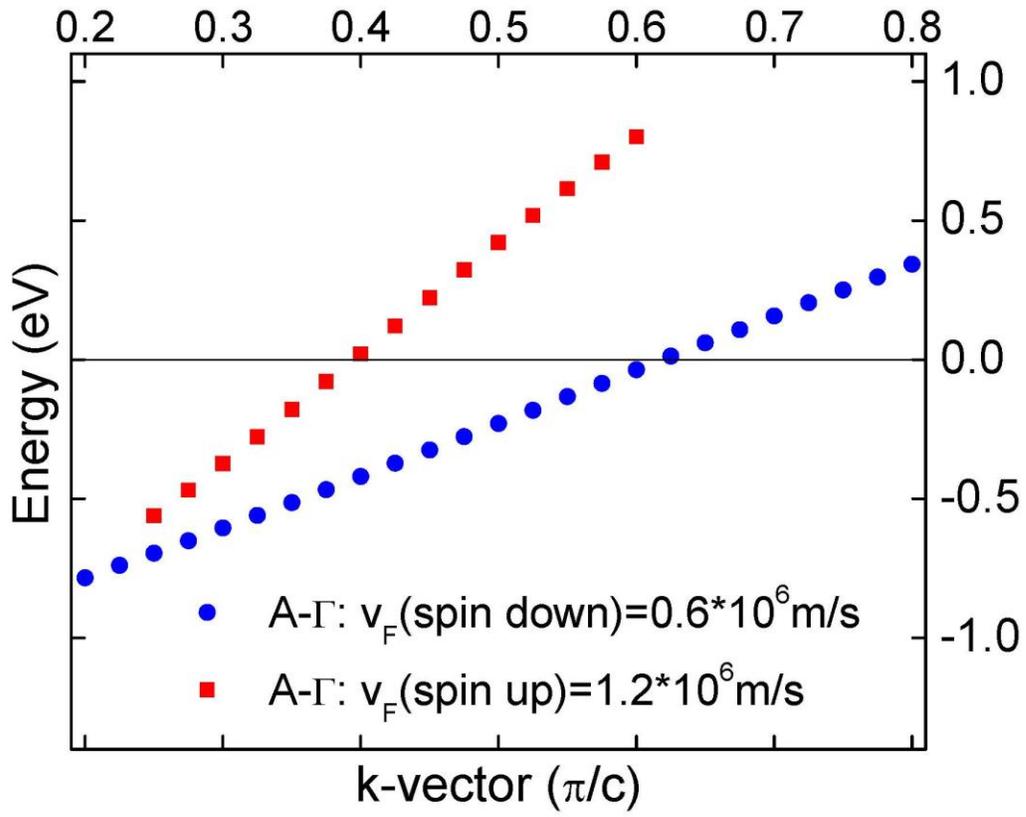

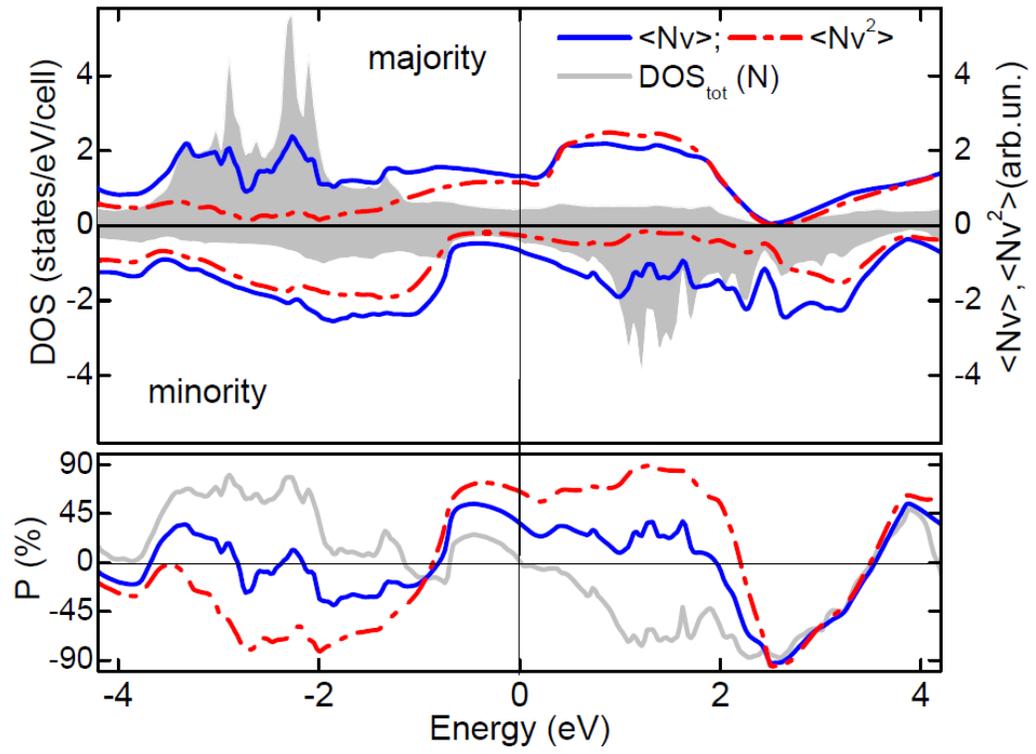